# Hierarchical Crossover and Probability Landscapes of Genetic Operators

Stefan Bornholdt[a,b] and Heinz Georg Schuster[a]

a) *Institut für Theoretische Physik, Universität Kiel*
*Leibnizstr. 15, D-24098 Kiel, Germany*

b) *Santa Fe Institute*
*1399 Hyde Park Road*
*Santa Fe, NM 87501, USA*

## Abstract

The time evolution of a simple model for crossover is discussed. A variant of this model with an improved exploration behavior in phase space is derived as a subset of standard one- and multi-point crossover operations. This model is solved analytically in the flat fitness case. Numerical simulations compare the way of phase space exploration of different genetic operators. In the case of a non-flat fitness landscape, numerical solutions of the evolution equations point out ways to estimate premature convergence.

During the last decade, genetic algorithms [1, 2] have advanced to very powerful optimization tools with real world applications in many different fields [3]. The theoretical understanding of these algorithms, however, has not kept pace with this development. After a brief overview of the different approaches to an understanding of genetic algorithms, we will present an alternative view from the perspective of statistical mechanics.

The basic mechanism of genetic algorithms works as follows. The set of parameters of a given problem is coded as an $N$ dimensional binary vector $\{x_i\}$ with $x_i \in \{0,1\}$ for all components $i$. Then the task is to find an optimal solution by looking for a maximum of a suitably defined fitness function in this $N$ dimensional parameter space. Basic mechanisms to scan this space are mutation and crossover, followed by a selection step where the fittest vectors are selected. Let us consider the simple case of a flat fitness landscape with fitness function $f = 1$. Mutation moves the vectors in this space by stochastic flips of single components of the parameter vectors. It covers the space by a process similar to diffusion spreading out via next neighbors into the search space. A completely different propagating behavior is exhibited by crossover. In its simplest form, it takes two random vectors and swaps a certain fraction of components between them. The simplest version is the one-point crossover, where all bits beyond a certain crossover point are exchanged. The produced "offspring" typically does not belong to the close neighborhood of the "parents" in the phase space, s.t. crossover is able to cover a large search space quite fast. Unlike mutation, it does not have to suffer from the strongly inhomogeneous nature of the diffusion process. Finally, it may be important to notice that if one starts with two maximally distant vectors, crossover is able to reach any other point in phase space.

The dynamics of mutation can be understood in terms of non-equilibrium statistical mechanics [4]. However, very little is known about the convergence and phase space dynamics of crossover. Due to the highly nonlinear nature of the crossover operator, a full calculation of the dynamics already without fitness quickly becomes complex, not to mention the evolution in arbitrary fitness landscapes where hardly anything quantitative can be said about the full time evolution. Another problem arises from the complicated dynamics of finite size populations. Different approaches have been taken to get a good understanding of how genetic algorithms work. Several limiting cases proved to be useful. The most general statements about the convergence of a genetic algorithm can be made in the limit of just one time step of the evolution. One finds inequalities about the change in frequency of the members in a population proving the convergence properties of genetic algorithms (Schema Theorem [1]), or estimates for the evolution of the mean fitness of a population (Price's Theorem [5]). A second approach is to explore the dynamics of a genetic algorithm for a *specific* fitness function. Functions have been studied that are considered to be particularly easy (royal road functions [6]) or hard (deceptive problems [7]) for a genetic algorithm. The time evolution of a genetic algorithm quickly becomes complicated, not at least due to the finite size of the populations. Approaches have been taken treating small populations as Markov chains [8, 9] on the one hand, and the infinite population limit of statistical mechanics [10, 11] on the other. Results of the latter approach have been shown to be of importance also to the dynamics of finite populations [9]. In the following we proceed along the lines of this limit of statistical mechanics.

In this article we study the time evolution of infinite population models under crossover. After deriving a one-point crossover model for a flat fitness landscape, we present a hierarchical model of crossover that is optimized for a fast and homogeneous exploration of the phase space. We solve the model analytically for a flat fitness surface and give results of numerical studies with flat fitness as well as a rugged fitness function of the travelling salesman problem. The "hierarchical crossover" operator proves to cover the phase space fast and – in the case of a flat fitness – homogeneously. Furthermore, the coverage occurs in a true hierarchical fashion.

First let us consider how mutation moves into phase space in the example of a simple spin chain. We start from a given vector $\{x_i\}$ with $x_i = 1$ for all $i$. In each time step let us flip one of the components into the opposite state. How long does it take to reach *any* vector in the phase space from one given starting vector? The minimum time is simply given by the number of required spin flips, $N$ in this case. However, the probability to reach a distant state scales badly with dimension $N$ due to the diffusion type dynamics of mutation [4].

One-point crossover exhibits a better scaling with dimension. Let us start with two maximally distant parent vectors $\{x_i\}$ with $x_i = 1$ and $\{y_j\}$ with $y_j = 0$ for all $i, j$. In every step, the operator generates a new "twist" location in the spin chain pair. In addition, it is able to put together two parts containing possible earlier twists. Therefore, the minimum time $t$ to reach *any* vector in the $N$ dimensional phase space is given by the condition $t \geq \log_2 N$. However, this optimum is not efficiently implemented in the standard one-point crossover. The reason is that the two vector parts put together by the crossover operator usually did not experience the maximum number of earlier crossover flips. Nevertheless, they remain in the gene pool and reach the target vector at a later time than the optimal combination of recombination steps. The idea of "hierarchical crossover" is to eliminate these sub-optimal paths of crossover evolution and just retain the shortest paths that lead to any target vector in the phase space.

To be more specific we first consider the dynamics of a two dimensional string under crossover in the limit of a large gene pool. The genes are strings of two binary variables $x_1$ and $x_2$ with values $x_{1,2} \in \{0, 1\}$. The probability to draw a specific string from the pool is given by $P^0(x_1, x_2)$. We are interested in the probability $P^t(x_1, x_2)$ to find the string at later times $t$. For a crossover probability $p$, we obtain after the first time step:

$$P^1(x_1, x_2) = (1-p)P^0(x_1, x_2) + pP^0(x_1)P^0(x_2) \tag{1}$$

where

$$P^0(x_1) = \sum_{x_2} P^0(x_1, x_2) \tag{2}$$

is the probability to find a string with a specified value of $x_1$. The partial probability $P^0(x_1)$ corresponds to the probability of a schema $(x_1, *)$ in the traditional formalism of genetic algorithms [1]. We can now define an operator $C^2$ which describes this decomposition of the probability of a state into partial probabilities through crossover within a population. Define $C^2$ by

$$\begin{aligned} \sum_{y_1, y_2} C^2 P^t(x_1, x_2) P^t(y_1, y_2) &= \sum_{y_1, y_2} P^t(y_1, x_2) P^t(x_1, y_2) \\ &= P^t(x_1) P^t(x_2) \end{aligned} \tag{3}$$

which gives the probability for a state $(x_1, x_2)$ to be produced by crossover in the time step $t \to t+1$. (The superscript 2 refers to breaking up the probability into two partial probabilities, the only way for $N = 2$). This accounts for the fact that the crossover operation within a population takes all possible pairs of strings and, as in (1) with probability $p$, exchanges the first components between them. One can now derive the second time step by writing

$$P^2(x_1, x_2) = \sum_{y_1, y_2} [(1-p)1 + pC^2][(1-p)P^0(x_1, x_2) + pP^0(x_1)P^0(x_2)]$$
$$\times [(1-p)P^0(y_1, y_2) + pP^0(y_1)P^0(y_2)]. \quad (4)$$

In general one obtains for $t \geq 2$

$$P^t(x_1, x_2) = (1-p)^t P^0(x_1, x_2) + [1 - (1-p)^t] P^0(x_1) P^0(x_2). \quad (5)$$

The distribution of strings in the gene pool at any time follows directly from the initial distribution. In general, for large $N$, crossover operates at different points of the strings and contributes $N - 1$ different terms at each time step. The expressions for the evolution with time become large quickly and a solution for general $N$ is not readily obtained.

Let us consider the recombination paths in one-point crossover. Choosing the crossover points at any position with equal probabilities we obtain for arbitrary $N$

$$P^{t+1}(x_1 \ldots x_N) = (1-p)P^t(x_1 \ldots x_N) + \frac{p}{N-1}[P^t(x_1)P^t(x_2, \ldots, x_N) +$$
$$P^t(x_1, x_2)P^t(x_3, \ldots, x_N) + \ldots + P^t(x_1, \ldots, x_{N-1})P^t(x_N)]. \quad (6)$$

After several time steps, a given state may have many different possible origins via the different possible combinations of crossover operations leading to the same state. For the case $N = 4$ this is shown in figure 1. For $p = 1$ the different paths of crossover are shown in terms of sub-string probabilities. Note that these paths in general have different lengths. In this case, the shortest path reaches any state after two time steps. In the following we will take a closer look on just this optimal path in the evolution. It is the marked path in the middle where the crossover point is always chosen in the middle of any yet "untouched" (sub)string. The other paths to the right and left take one step longer. In principle they are redundant since the path in the middle is not only sufficient but even more economical. Below we will find that this branch leads to an appealing analytical form of the evolution equation. Furthermore we will pursue the idea to construct a crossover operator which omits the redundant terms. In other words, in each step we choose only the most effective crossover points from the repertoire of standard crossover.

The dynamics are hard to depict, especially in a high dimensional phase space. For $N = 4$, the neighborhood relations between different states are simple enough such that the basic idea can be seen in a two dimensional picture. This is shown in figure 2. Here, the phase space is shown in the second time step after starting with the initial states (0000) and (1111) at $t = 0$. Mutation only proceeds to next neighbors in each time step, here depicted as horizontal bars. States with many bits differing from the initial state are reached only in later steps. Crossover is able to scan phase space beyond next neighbors and reaches all states on the circle in the next picture. This skipping nature is known as the special feature of crossover. The third frame shows how hierarchical crossover scans phase space. It reaches the boxed states at $t = 1$ which is the maximally distant pair of states to the initial pair. It retains the feature of crossover omitting the overhead of redundant states at early times. In the analytical formulation, hierarchical crossover operates on strings of length $N = 2^n$, $n$ being an integer. For $N = 4$ we obtain at $t = 1$

$$\begin{aligned} P^1(x_1, x_2, x_3, x_4) &= \sum_{y_1, y_2, y_3, y_4} \left[(1-p)1 + pC^2\right] P^0(x_1, x_2, x_3, x_4) P^0(y_1, y_2, y_3, y_4) \\ &= (1-p) P^0(x_1, x_2, x_3, x_4) + p P^0(x_1, x_2) P^0(x_3, x_4), \end{aligned} \quad (7)$$

where $C^2$ chooses the crossover point in the middle of the strings. All total probabilities are normalized to 1. The next step is

$$\begin{aligned} P^2(x_1, x_2, x_3, x_4) &= \sum_{y_1, y_2, y_3, y_4} \left[(1-p)1 + pC^4\right] \\ &\times [(1-p) P^0(x_1, x_2, x_3, x_4) + p P^0(x_1, x_2) P^0(x_3, x_4)] \\ &\times [(1-p) P^0(y_1, y_2, y_3, y_4) + p P^0(y_1, y_2) P^0(y_3, y_4)]. \quad (8) \end{aligned}$$

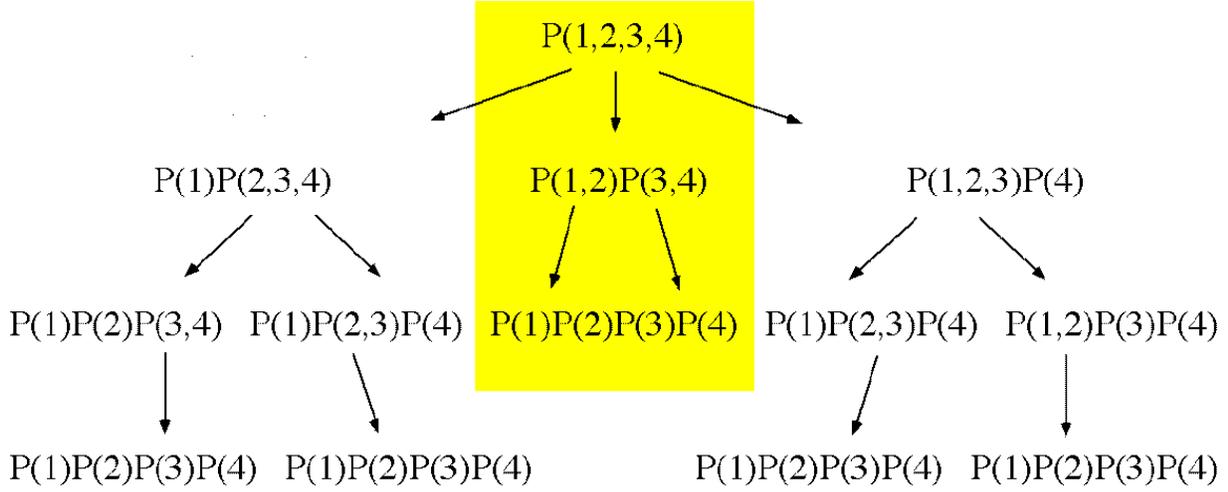

Figure 1: Different crossover paths for $N = 4$

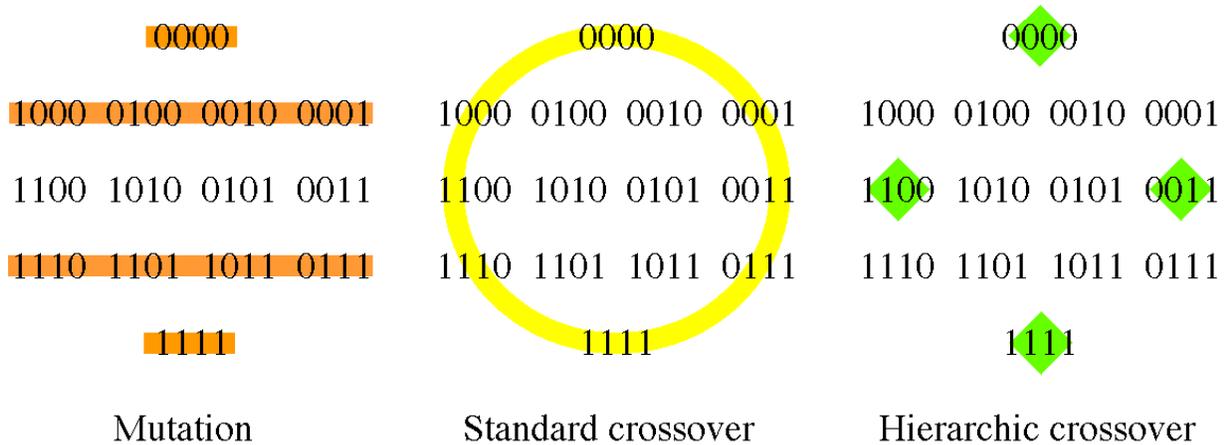

Figure 2: Phase space exploration at $t = 1$ for initial states (0000) and (1111) at $t = 0$

Here, the operator $C^4$ swaps the $x_1$ and $x_3$ components such that one obtains the particularly simple form

$$P^2(x_1, x_2, x_3, x_4) = (1-p)^2 P_4 + (1-p)p P_2 + p P_1. \tag{9}$$

where we denote

$$P_4 = P^0(x_1, x_2, x_3, x_4) \tag{10}$$
$$P_2 = P^0(x_1, x_2) P^0(x_3, x_4) \tag{11}$$
$$P_1 = P^0(x_1) P^0(x_2) P^0(x_3) P^0(x_4). \tag{12}$$

Furthermore, we assume that the operator $C^4$ always returns an expression of type $P_1$, no matter which combination of $P_2$ and $P_4$ type expressions it operates on. Any further application of this crossover operator essentially increases the $P_1$ term by decomposing more of the $P_2$ and $P_4$ terms, while acting as an identity on the pure $P_1$ states such that, for $t \geq 2$, $P^t(x_1, x_2, x_3, x_4)$ is given by

$$P_4^t = (1-p)^{t-1}[(1-p)P_4 + pP_2] + [1 - (1-p)^{t-1}]P_1. \tag{13}$$

One can generalize this formalism to arbitrary dimensions $N = 2^n$ and obtains for $t \geq n$

$$P_{2^n}^t = (1-p)^{t-n+1}\left[(1-p)^{n-1} P_{2^n} + (1-p)^{n-2} p P_{2^{n-1}} + \ldots + p P_2\right]$$
$$+ \left[1 - (1-p)^{t-n+1}\right] P_1. \tag{14}$$

The main feature of this result is that after only $n$ crossover steps the $n$th step produces *all* possible states (with *equal* probabilities if $p = 1$). The phase space exploration occurs in a hierarchical fashion using only the shortest possible path for each state. Any further evolution increases the density of this distribution.

In order to obtain this behavior in practical simulations one chooses a chromosome of length $2^n$. Furthermore, this formalism uses a modified crossover operator. The construction of this operator will be described in the following. In (9) when describing the evolution of a whole population it is easy to guess the result of any further application of the crossover operator $C^4$: The result is always proportional to $P_1$. In the case $p = 1$ we simply have to apply the crossover operators $C^{2^{t+1}}$ one after the other starting with $C^2$ at $t = 0$ until one reaches $C^n$ for strings of the length $N = 2^n$. If $p \neq 1$, we have to be slightly more careful. Now, not the overall time $t$ determines which operator we have to take, but rather the number of crossover operations that the individual strings have undergone so far. We introduce an individual "age tag" to every individual, denoting by which operator it has been produced. A string produced by $C^2$ has age 1, one from $C^4$ age 2 etc. The prescription for the crossover procedure for one time step within a population is then the following:

- For each pair in the population determine the minimum age $a$.

- Apply the crossover operator $C^{2^{a+1}}$ to each pair. The children are assigned the age $a' = a + 1$, where $a$ is the smallest age tag of the two parents.

- The children with $a' = t + 1$ are transferred to the next generations gene pool.

- The children with $a' < t+1$ have to be processed further. Build all possible pairs from *all* children and group the pairs in different *sub-populations* of the same age tags of the pairs, e.g., pairs with ages (1,1), (1,2), (2,1), etc., except for pairs of the type (t+1,t+1).
- Perform successive crossover *within* each sub-population until the children have all age $t+1$.
- Mix the members within each subpopulation between successive crossover steps.
- Once the members of all sub-populations have reached age $t+1$, add them to the next generations gene pool.

Some remarks are due concerning this modified crossover. First of all, when expressing this prescription in the earlier described formalism, one can show that it indeed corresponds to the desired behavior of the hierarchical crossover operator in (14). Furthermore, in the limit of large populations (which is the limit of the analytical equations), some of the operations within the sub-populations are just mimicking earlier operations, so in this limit, the procedure can be simplified further.

In the following we present this algorithm in numerical simulations and compare it to mutation and one-point crossover, first, for a flat fitness surface. The simulations follow the complete evolution of the phase space, starting from two maximally distant vectors $(0,0,...,0)$ and $(1,1,...,1)$. This simulation corresponds to the limit of very large populations in a regular genetic algorithm. This limit has been proven useful earlier to describe the average behavior of genetic algorithms with large populations [9].

In figure 3 the filling of the phase space for a small $N=8$ model is shown as the probabilities of all states, forming a "probability landscape" over the whole space. The leftmost squares shows the initial condition which is the same for all simulations with the two maximally distant vectors. The 256 states of the phase space are depicted in a 16 by 16 field where the coordinates follow the decimal values of the leading 4 and least 4 bits of the string, with the states coded as 0 or 1. I.e., the lower left corner of each square corresponds to (00000000), the upper right one to (11111111), and the upper left to (11110000) and the lower right lower to (00001111). The maximum value on the Z axis corresponds to a probability of a state to occur of 0.5 (red) on a logarithmic scale to small values (blue), with 0 being black.

The upper row of figure 3 shows the evolution under mutation with mutation probability 0.5. This diffusion-like process fills the space slowly beginning near the starting points. The exact diffusion type nature would be visible in an 8 dimensional picture, similar to mutation in figure 2. The probability density is very inhomogeneous. The one-point crossover (with p=1) in the middle row of figure 3 performs better. Due to the recombination, also distant points like the ones at the orthogonal corners are reached already at $t=2$. The probability distribution is inhomogeneous by several orders of magnitude. In the lower part of figure 3 the evolution for hierarchic crossover is shown (for $p=1$). The phase space is covered after 3 time steps and the probability distribution of the states produced in step 3 is homogeneous (for all states if $p=1$). This homogeneous covering of phase space was obtained by just omitting irrelevant operations from standard crossover procedures!

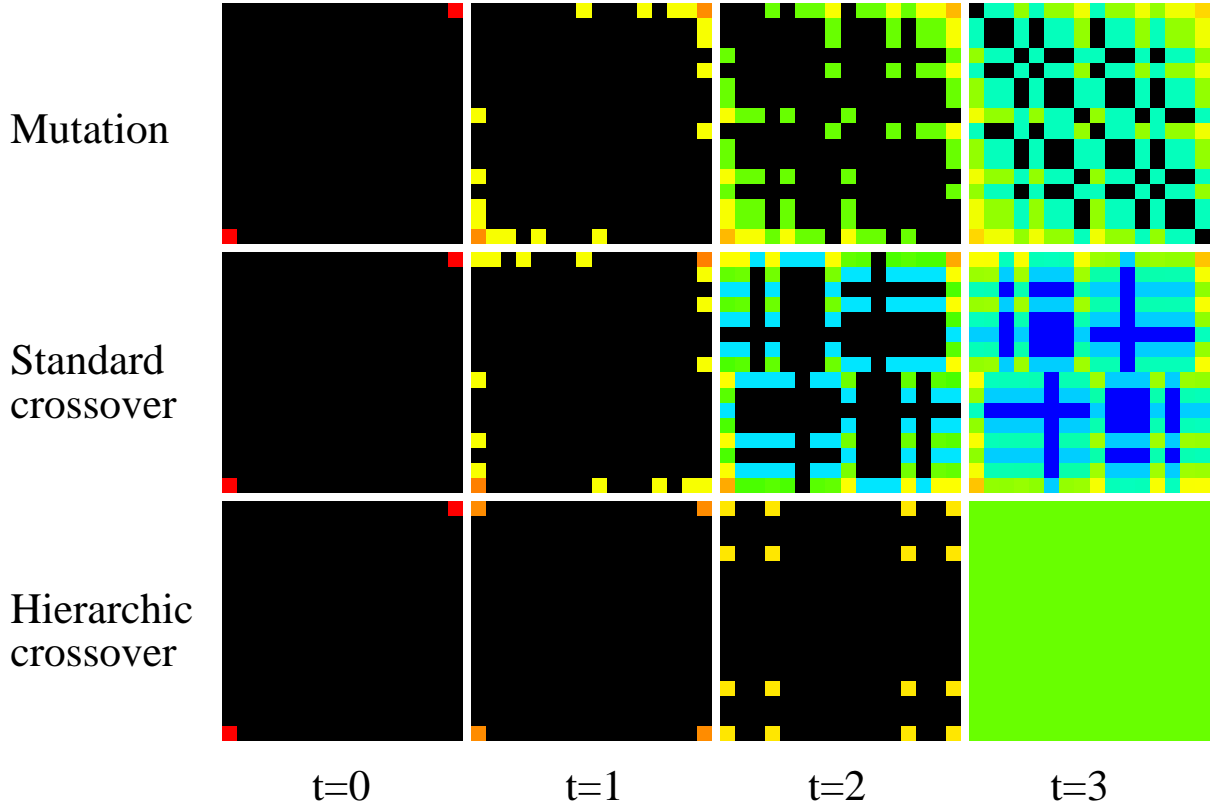

Figure 3: Evolution of $N = 8$ states for times $t = 0, \ldots, 3$

What can we derive from the structure of these "probability landscapes" of genetic operators for practical problems with a non-flat fitness? In that case the "probability landscape" of the genetic operator interacts in a non-trivial way with the fitness landscape of the given problem. In order to investigate this, we will in the last part of this article apply this formalism to a problem with a more rugged fitness. For this purpose, we implement a small travelling salesman problem on an $N = 8$ string and follow the complete evolution of the phase space. We choose 5 cities with one kept fixed and code the position on the tour of the remaining 4 cities in an 8 bit string. The tour length $L$ is taken to be

$$L = \frac{1}{2} \sum_{i,j,k} D_{ij} I_{i,k} (I_{j,k+1} + I_{j,k-1}) \qquad (15)$$

with $I_{i,k} = 1$ if city $i$ is $k$th on the tour, else 0, and $D_{ij}$ being the distance between cities $i$ and $j$. The energy function $H$ contains the length plus penalty terms for multiple occurrence of cities and for preferring one direction

$$H = L + \frac{1}{2} \sum_{i} (1 - \sum_{k} I_{ik})^2 + \frac{1}{4} \theta \qquad (16)$$

with $\theta = 1$ if the number of the first city is larger than the number of the last city, or else $\theta = 0$. The fitness function $f$ is chosen as

$$f = e^{-\beta H} \qquad (17)$$

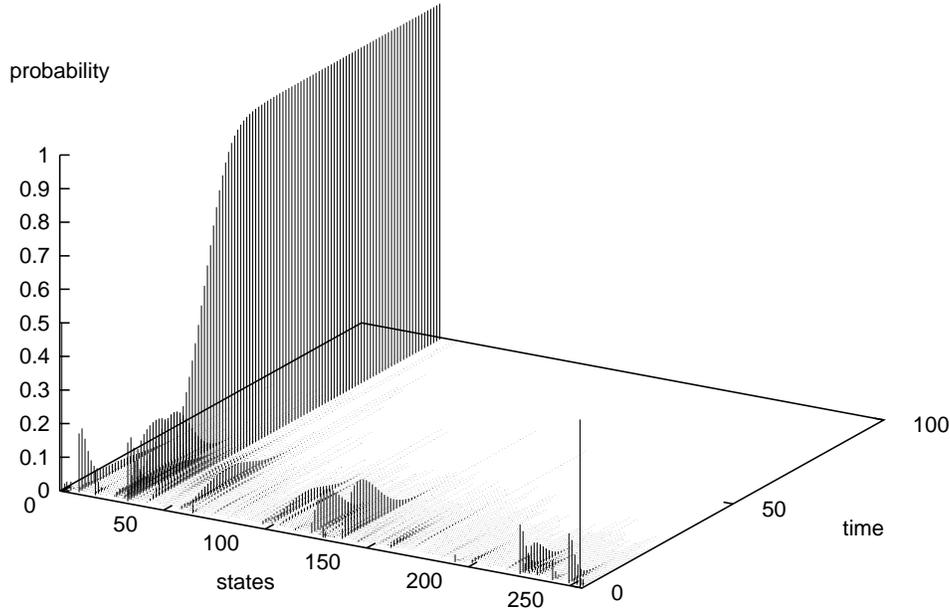

Figure 4: Travelling salesman with 5 cities under standard crossover, $\beta = 0.1$

with a free parameter $\beta$ to adjust its "ruggedness". The evolution equations have to be modified to contain fitness. In particular, this requires a normalization to preserve the probabilities. The above evolution equations are to be multiplied by the fitness $f$ and then normalized. For the simplest case $N = 2$ in the first time step this is

$$P^1(x_1, x_2) = \frac{f(x_1, x_2)[(1-p)P^0(x_1, x_2) + pP^0(x_1)P^0(x_2)]}{\sum_{y_1, y_2} f(y_1, y_2)[(1-p)P^0(y_1, y_2) + pP^0(y_1)P^0(y_2)]} \qquad (18)$$

and accordingly for the higher terms. For standard one-point crossover we used for the simulation the evolution equation

$$P^{t+1}(x) = \frac{f(x)[(1-p)P^t(x) + \frac{p}{N-1}\sum_{k=1}^{N-1} P_k^t(x)]}{\sum_y f(y)[(1-p)P^t(y) + \frac{p}{N-1}\sum_{k=1}^{N-1} P_k^t(y)]} \qquad (19)$$

where $P(x)_k$ denotes $P(x_1, ..., x_k)P(x_{k+1}, ..., x_N)$.

A simulation of the full evolution under these equations is more than the run of a genetic algorithm: It determines the expected average evolution of a genetic algorithm with a large population. In figures 4 and 5 the results of the simulations are shown. Figure 4 shows how standard one-point crossover solves the problem. The fitness function of the travelling salesman problem has been chosen s.t. the fitness optimum lies in one of the less dense regions of the probability landscape of one-point crossover. The simulations show that in this case the coincidence of a rugged fitness landscape of the problem with a rugged probability landscape of the genetic operator, sharply increases the probability of premature convergence towards a false minimum of the energy function. Only for a very smooth parametrization of the fitness landscape ($\beta \leq 0.1$) the evolution converged towards the right solution. In that case, it takes about 20 generations for the right solution to appear. Larger values of $\beta$ like 1 or 10 force premature convergence into false maxima. This results from the very inhomogeneous probabilities of the states produced by crossover. In figure 5 one can see that hierarchic crossover solves the problem even for very inhomogeneous fitness landscape ($\beta = 10$) already in the third time step. It turns out to be very robust against steep fitness functions.

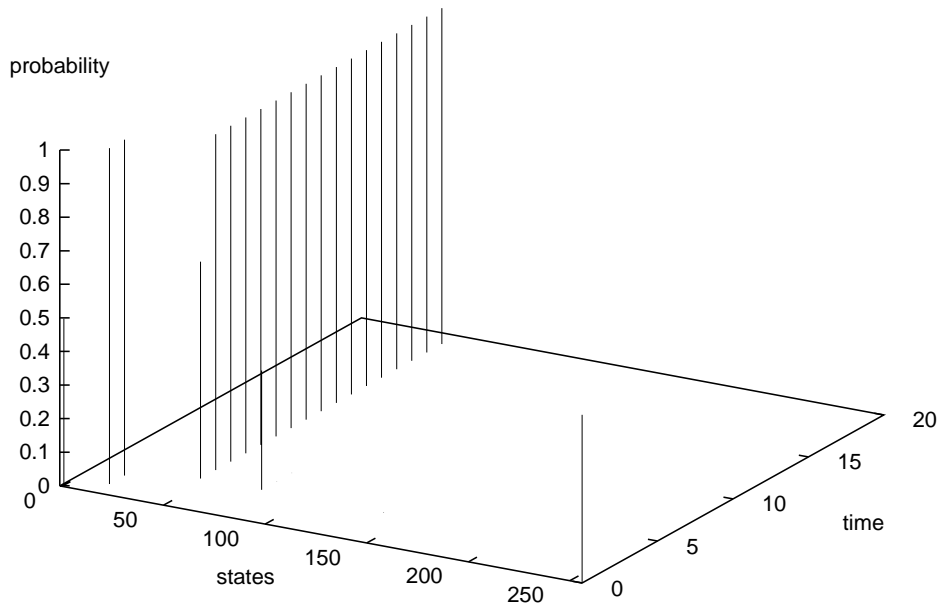

Figure 5: Travelling salesman, 5 cities, under hierarchic crossover, $\beta = 10$

In this article, we derived a simple model for crossover. The discussion of the time evolution of this model (over more than one generation) lead us to a simple variant with an improved exploration behavior in phase space. This model is analytically solvable in the flat fitness case. We developed the concept of the "probability landscape" of a genetic operator as opposed to the "fitness landscape" of the underlying problem. Numerical simulations suggest that genetic operators with a homogeneous probability landscape are more robust against premature convergence. Inhomogeneities in both, probability *and* fitness landscapes, appear to favor premature convergence. The next steps in this investigation include the generalization of the formalism to the case with nontrivial fitness functions. Although a hopeless task for general models of crossover, this might be more feasible for the model of hierarchical crossover, at least in some special situations as, e.g., in the statistical case of random fitnesses. Furthermore, it has to be explored, how the advantages of the hierarchical crossover operator translate to finite size populations and large search spaces. The goal of this approach is a better understanding of the dynamics of genetic algorithms.

## Acknowledgements

We wish to thank Raja Das, Stephanie Forrest, Terry Jones, Melanie Mitchell, Klaus Schmoltzi, and Derek Smith for their critical comments and helpful discussions with respect to an earlier version of this paper. This work has been supported by the Deutsche Forschungsgemeinschaft and the Santa Fe Institute.